\documentclass[a4paper,11pt]{article}
\pdfoutput=1 
\usepackage{jinstpub} 
\usepackage{subfig}

\usepackage{siunitx}
\usepackage[capitalise,nameinlink]{cleveref}
\usepackage{lineno}

\title{\boldmath The new monolithic ASIC of the preshower detector for di-photon measurements in the FASER experiment at CERN}
\author[a]{S. Gonzalez-Sevilla}
\affiliation[a]{D\'partement de Physique Nucl\`eaire et Corpusculaire (DPNC), University of Geneva,\\CH-1211, Geneva, Switzerland}
\emailAdd{Sergio.Gonzalez@unige.ch}

\abstract{The ForwArd Search ExpeRiment (FASER) is an experiment searching for new light and weakly-interacting particles at CERN's Large Hadron Collider. FASER is composed of different sub-detectors, including silicon microstrip detectors, scintillator counters and an electromagnetic calorimeter. In this paper, a new preshower detector for the FASER experiment is presented. The new detector, based on monolithic pixel ASICs, will provide excellent spatial and time resolutions and a large charge dynamic range. First results from a prototype chip produced by IHP in 130 nm SiGe BiCMOS technology are shown.}


\collaboration[c]{(for the FASER Collaboration)}

\usepackage{xspace}
\newcommand{\xz}{\ensuremath{\text{X$_0$}}\xspace}

\proceeding{23$^{\text{rd}}$ International Workshop on Radiation Imaging Detectors (iWoRiD)\\
26-30 June 2022\\
Riva del Garda, Italy}

\begin{document}
\maketitle
\flushbottom

\section{Introduction}
\label{sec:intro}

The quest for New Physics by the large experiments at the Large Hadron Collider (LHC) is typically based on signatures from heavy and strongly interacting particles ({\em e.g.} high transverse momentum $\rm{p_T}$, large missing transverse energy). FASER (ForwArd Search ExpeRiment)~\cite{FASER} adopts an alternative approach and will search for light (MeV to GeV range) and weakly-interacting new particles (long-lived particles, LLPs), dominantly produced at low $\rm{p_T}$ from rare meson decays~\cite{FASER:2018eoc}. As these new particles are very weakly coupled, large production rates are needed to enhance their detection. FASER will exploit the large total inelastic proton proton cross section at the LHC ($\sim \SI{75}{\milli\barn}$ at 13 TeV), where $\sim 10^{16}$ inelastic events are expected to be produced in Run 3 ($150\,\rm{fb^{-1}}$), mostly in the forward direction. FASER is located $\sim\SI{480}{\metre}$ downstream the ATLAS interaction point (IP1) along the beam collision axis line-of-sight (LOS) in the unused service tunnel TI12. A LLP produced at IP1 (e.g. dark photon $A'$ produced from the rare decay of $\pi^0$ or $\eta$ mesons) may travel in the very forward direction along the LOS \mbox{before decaying to visible particles in the decay volume of the detector.}

\begin{figure}[t]
\centering
\includegraphics[width=\textwidth]{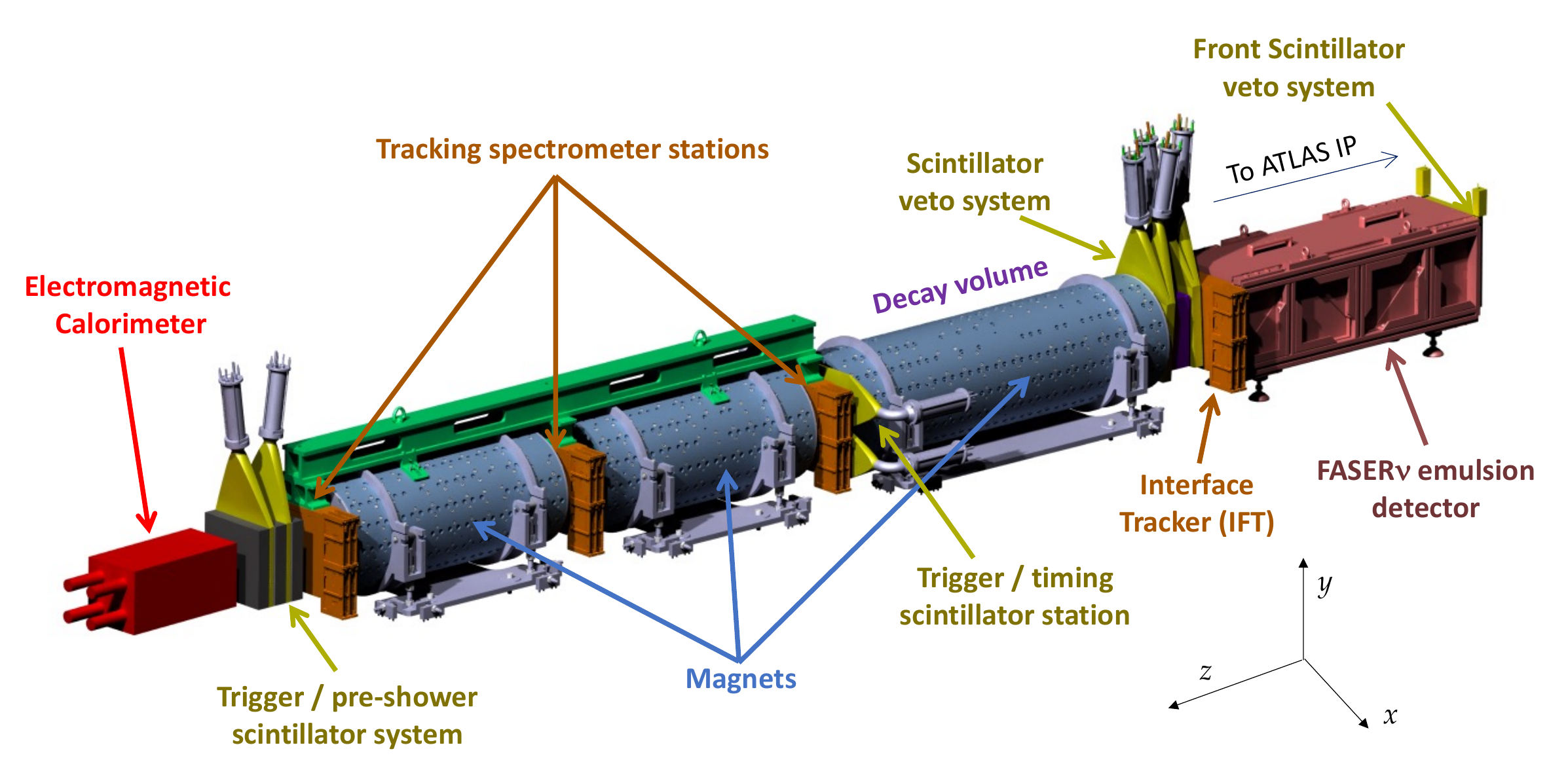}
\caption{\label{f:faserDetector}The FASER detector. Particles produced from the collisions at the interaction point 1 of the LHC enter the detector from the right. The aperture radius of the magnets is 10 cm and the total length of the experiment is $\sim$ \SI{7}{\metre}. The sensitive area in the transverse direction ($x-y$ plane) is larger than the magnets' aperture, approximately $30\times 30\,\rm{cm^2}$ for the tracking layers and the veto and preshower stations, and $40\times 40\,\rm{cm^2}$ for the timing station.}
\end{figure}

A sketch of the experiment is shown in \cref{f:faserDetector}. The main detector consists of a magnet system, three tracking stations, three scintillator stations and a calorimeter~\cite{FASER:2022hcn}. It is complemented with the FASER$\nu$~\cite{FASER:2020gpr} emulsion detector (to study neutrino interactions), a fourth tracking station (Interface Tracker, IFT) and an additional upstream scintillator station. The magnet system comprises three permanent dipole magnets with a \SI{20}{\centi\metre}-diameter aperture each providing a field of $\sim 0.57\,\rm{T}$. A first upstream \SI{1.5}{\metre}-long magnet serves as decay volume to the LLPs. The tracking spectrometer~\cite{FASER:2021ljd} is formed by the other two (\SI{1}{\metre}-long) magnets interleaved between three tracking stations. Each tracking station is composed of three planes of eight double-sided silicon microstrip detectors, spares from the ATLAS Semicondutor Tracker (SCT)~\cite{Jackson:2005ec} barrel module production. The electromagnetic (EM) calorimeter consists of four spare modules from the LHCb outer EM calorimeter~\cite{Machikhiliyan:2009zz}. The first two scintillator stations are used to veto with high efficiency ($>99.9\%$) any incoming charged particle, typically high energy muons produced at IP1. The third scintillator station placed in front of the spectrometer provides a trigger signal for charged particles emerging from the decay volume (e.g. $A'\rightarrow e^+e^-$). Finally, the last scintillator station, consisting of two scintillator counters interleaved with two tungsten absorber plates, provides an additional trigger signal in coincidence with the the first trigger station and acts as a simple preshower for the EM calorimeter. All scintillator counters are optically coupled to  photomultiplier tubes. The scintillator and calorimeter signals are digitized and sent to a custom FPGA-based board (GPIO~\cite{FASER:2021cpr}) that combines them using look-up tables and generates a global trigger signal (L1A) in case of signals above pre-defined thresholds. The L1A trigger is sent to the different detector readout boards to retrieve the data. The expected trigger rate is $\sim \SI{650}{\hertz}$ at $2\times 10^{34}\,\rm{cm^{-2}\,s^{-1}}$, dominated by muons produced at IP1 collisions~\cite{FASER:2021cpr}.
 
The main FASER detector was installed in the TI12 tunnel in March 2021. It was succesfully commissioned in-situ with cosmic rays ($\sim 125\,\rm{M}$ events recorded) and with LHC pilot beam test runs in October 2021. The front veto scintillator station, the IFT tracking station and approximately 1/3$^{\text{rd}}$ of the FASER$\nu$ emulsion detector were installed few months after, in time to start the full detector commissioning with the first LHC Run 3 beams in spring 2022. FASER is now regularly taking data at the nominal LHC beam energy of 6.8 TeV. 

\section{A new preshower detector}
\label{sec:preshower}

\begin{figure}[t]
\centering
\includegraphics[width=0.95\textwidth]{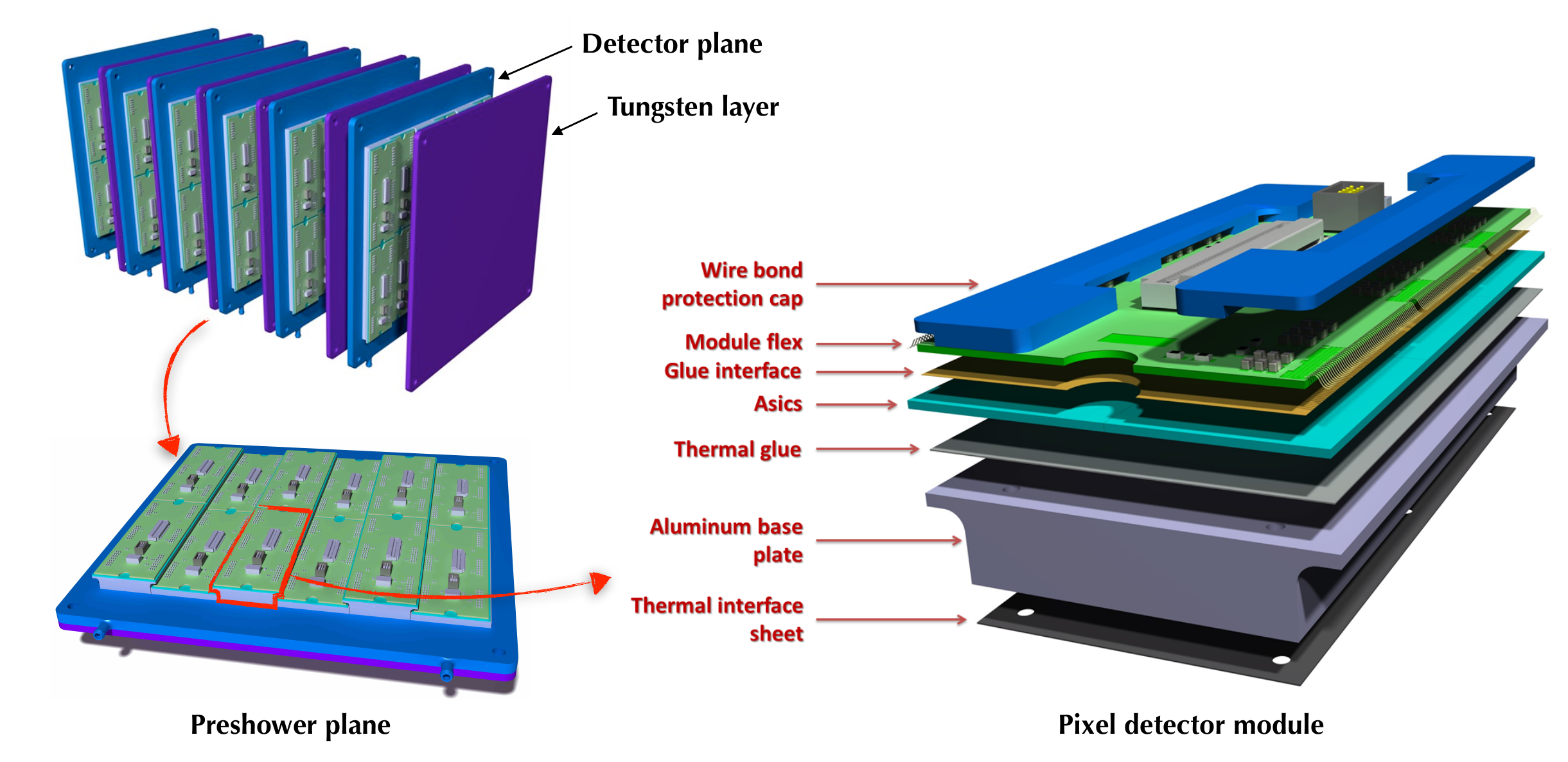}
\caption{\label{f:planeModule} Layout of the new preshower. On a detector plane the modules are staggered with an overlap of \SI{2}{\milli\metre} to minimize the effect of the dead area of the chip periphery.}
\end{figure}

The role of the current preshower station in FASER is to distinguish between signals coming from deep inelastic scattering of high energy neutrinos in the calorimeter and photons. The calorimeter modules have a Shashlik layout, with wavelength shifting fibers running along the module length perpendicularly to the absorber plates. Due to the lack of longitudinal segmentation of the sampling modules, two close-by high energy photons (e.g. coming from the decay of an \mbox{axion-like particle, ALP}) can't be resolved. For example, in the case of a light ($m\sim 100\,\rm{MeV}$) and highly boosted ($E\sim 1\,\rm{TeV}$) ALP decaying to a photon pair, the separation between the two photons is less than one millimetre. To enhance the experiment's sensitivity to the di-photon final state, a new high-granularity preshower detector based on monolithic pixel sensors has been recently proposed~\cite{WSiTp}. The baseline layout of the new preshower is shown in \cref{f:planeModule}. It consists of six pixel detector planes interleaved with \SI{3.5}{\milli\metre}-thick ($\sim$ 1 \xz) tungsten absorber layers. This layout (6 \xz) ensures a photon conversion probability $>99\%$ for events with one or two photons, and $>95\%$ in the case of events with four photons. Each pixel plane is made of $6\times 2$ detector modules mounted on a $\sim 20\times 20\,\si{\centi\metre\squared}$ aluminum cooling plate. A pixel module consists of six monolithic pixel ASICs glued to an aluminum base-plate. On top of the ASICs, a flex circuit implements the SMD components, connectors and wire-bonds required to drive power and control signals. The preshower layout has been optimized given the space constraints imposed by the current FASER detector. In particular, the longitudinal gap between the last tracking station and the EM calorimeter is only \SI{280}{\milli\metre}, with the additional requirement of keeping one scintillator layer for triggering purposes.

\section{The SiGe monolithic ASIC}
\label{sec:chip}

\begin{figure}[htbp]
\centering
\includegraphics[width=0.8\textwidth]{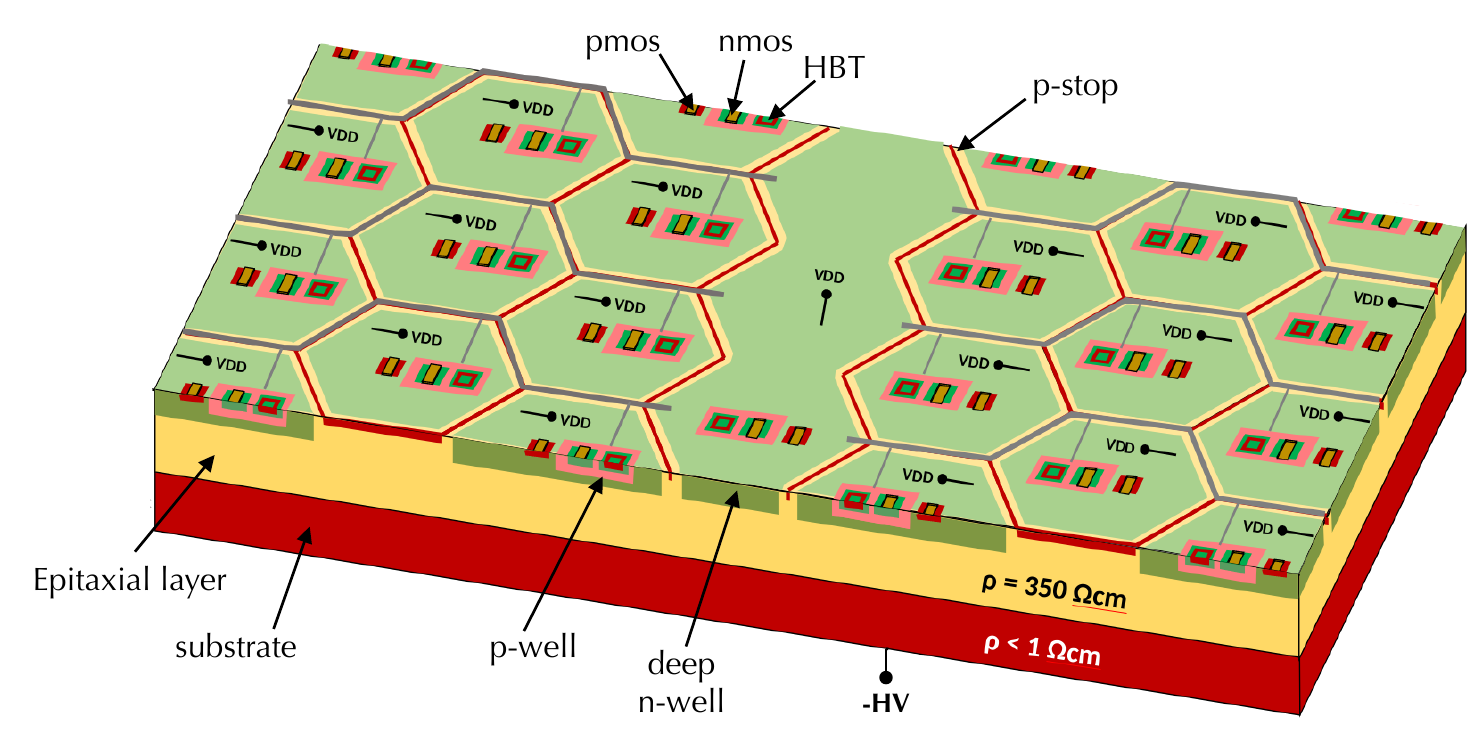}
\caption{\label{f:sensor} Sketch of the SiGe BiCMOS chip for the new FASER preshower. An example of die with a high resistivity (\SI{350}{\ohm\centi\metre}) p-type epitaxial layer on top of a low resistivity (\SI{1}{\ohm\centi\metre}) $\rm{p^+}$-type substrate is shown.}
\end{figure}

The proposed ASIC for the new FASER preshower modules is a monolithic pixel chip designed in 130 nm Silicon-Germanium (SiGe) BiCMOS technology (SG13G2 process by IHP Microelectronics~\cite{ihp}). The IHP BiCMOS technology combines heterojunction bipolar transistors (HBT) with CMOS transistors on the same die using a triple well structure. The SiGe BiCMOS technology provides fast signal amplification with low noise, resulting in excellent time resolutions~\cite{Iacobucci:2021ukp}. \Cref{f:sensor} shows a sketch of the chip structure. The pixels have an hexagonal shape (\SI{65}{\micro\metre} side) for a smoother distribution of the electric field lines on the edges. The pixel capacitance is \mbox{80 fF} and the inter-pixel pitch \mbox{is $\sim \SI{100}{\micro\metre}$}. A deep n-well implanted on a p-type substrate acts as collecting electrode. A shallow p-well embedded in the deep n-well isolates the NMOS and HBT transistors from the substrate that is set to a negative bias voltage. 

\begin{figure}[htbp]
\centering
\includegraphics[width=0.95\textwidth]{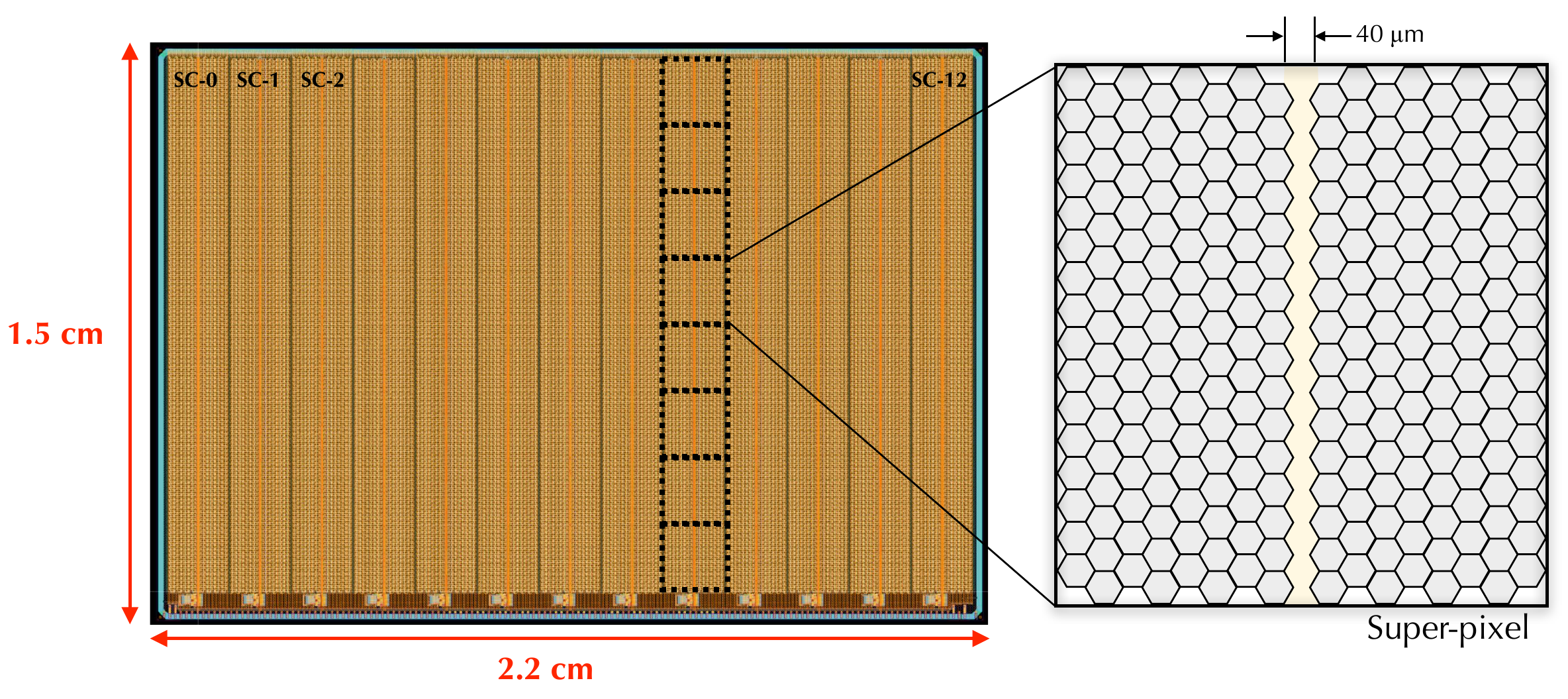}
\caption{\label{f:floorplan} Floorplan of the FASER chip. The digital periphery is at the bottom of the chip. The sketch on the right shows the arrangement of the $16\times 16$ pixels inside a super-pixel.}
\end{figure}

The floorplan of the preshower chip is shown in \cref{f:floorplan}. The $\sim 2.2\times 1.5\,\rm{cm^2}$ ASIC contains a matrix of 26'624 hexagonal pixels arranged in thirteen {\em super-columns} of 2048 pixels each. Each super-column is composed of an active region and a \SI{40}{\micro\metre}-wide digital column in the middle. The digital column is in a separated n-well isolated from the surrounding pixels via a p-stop guard-ring. The pixels inside a super-column are arranged in so-called {\em super-pixels}, with a super-pixel containing $16\times 16$ pixels, i.e. a single super-column consists of eight super-pixels.  The $\sim \SI{1}{\milli\metre}$-wide chip periphery includes the main digital circuits (TDC, arbitratrion logic, slow control), a seven guard-ring structure and the input/output pads. The main specifications of the chip are a large charge dynamic range from 0.5 fC to 65 fC, a maximum readout time of \SI{200}{\micro\second}, a power consumption less than \SI{150}{\milli\watt\per\centi\metre\squared} and a time resolution better than \SI{300}{\pico\second}. 

\begin{figure}[htbp]
\centering
\includegraphics[width=\textwidth]{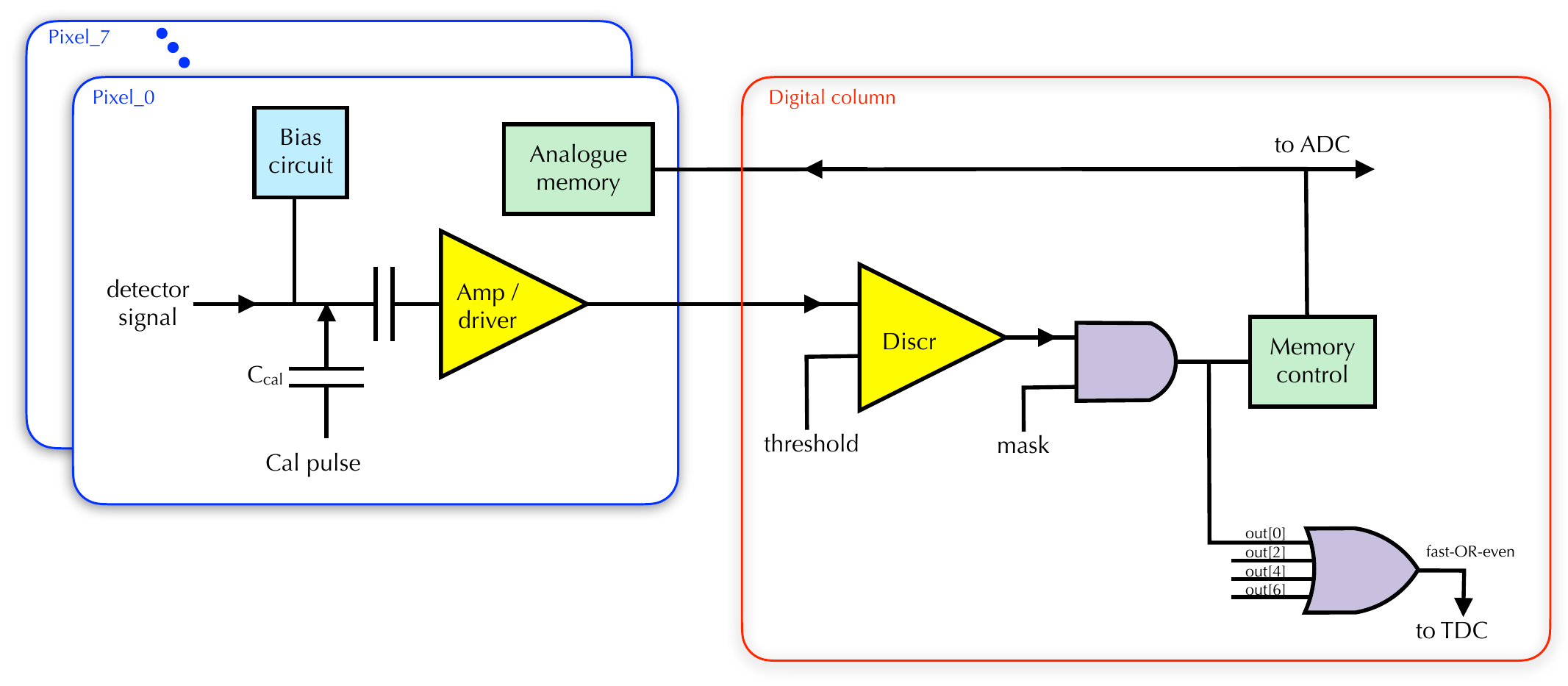}
\caption{\label{f:pixel}Conceptual diagram of a pixel row. For simplicity, only one side of the row is shown.}
\end{figure}

A pixel row corresponds to 16 pixels, i.e. eight pixels from each side of the digital column. \Cref{f:pixel} shows the basic structure of one side of the row. Each pixel contains the preamplifier, the bias circuitry and an analogue memory. A test-pulse of selectable amplitude can be injected directly to the preamplifier via a \SI{55}{\femto\farad} calibration capacitor. Due to the large dynamic range, the preamplifier has been designed to produce a signal (time-over-threshold, ToT) proportional to the logarithm of the input charge. The amplifier output is connected to a comparator implemented in the digital column, and while the signal is above threshold, the memory control connects the pixel analogue memory to a load current to produce a linear charge (corresponding to the ToT). Each pixel can be masked after the discrimination stage. 

Inside a super-pixel the pixels are arranged in three different groups, with a fast-OR output signal per group connected to the digital periphery. At the periphery, the time of arrival of the fast-OR signals is digitized by a multi-channel Time-to-Digital Converter (TDC) and a trigger signal is generated from the arbitration logic to initiate the chip readout. At this stage, the charge stored in the analogue memories is digitized with a 4-bit flash-ADC. The data is then output on a single Low Voltage Differential Signal line at 200 Mbps toward the backend acquisition boards.

\section{The pre-production chip and first results}
\label{sec:results}

\begin{figure}[ht]
\centering
\subfloat[\label{f:ret}]{\includegraphics[width=0.62\textwidth]{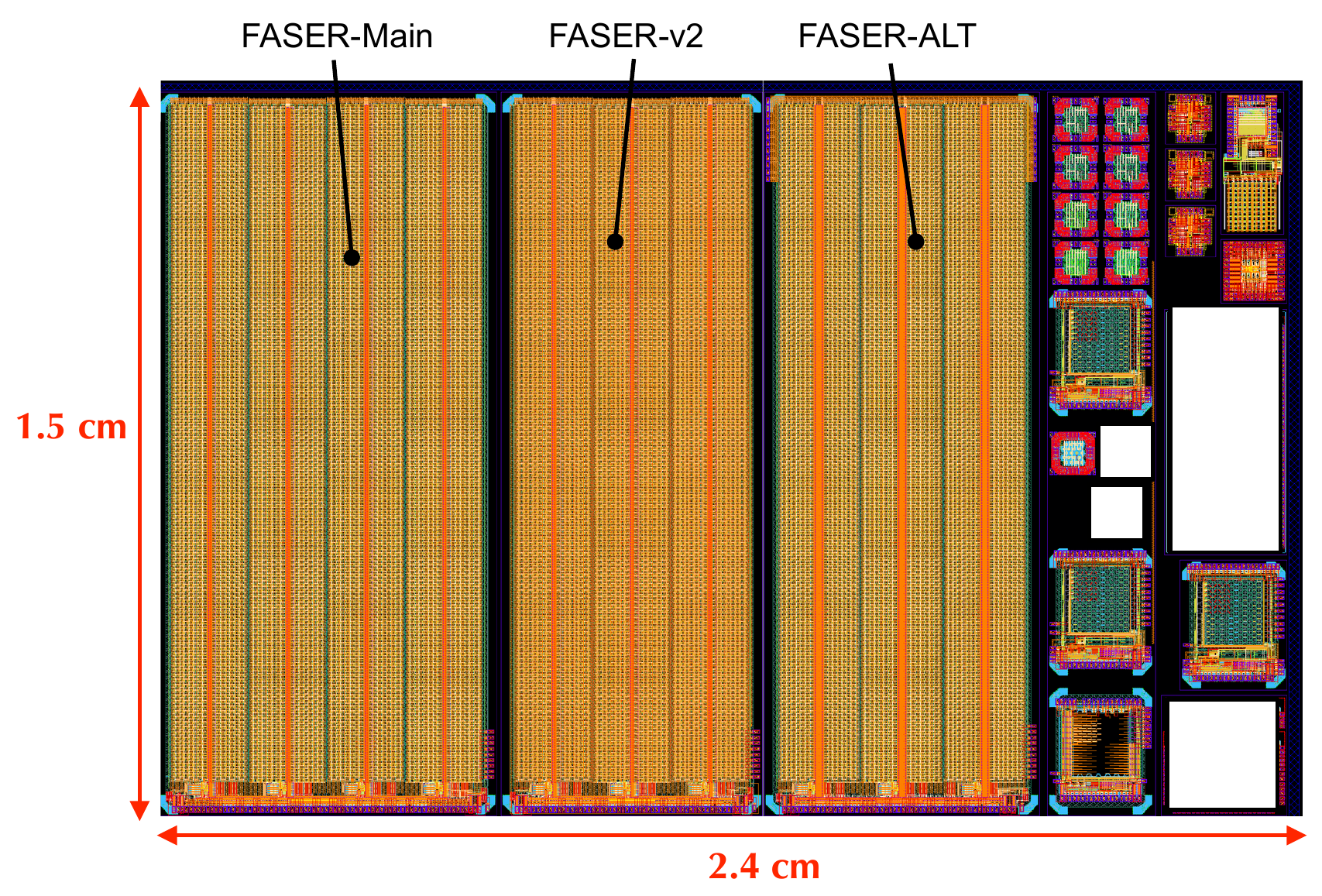}}\hfil
\subfloat[\label{f:waf}]{\includegraphics[width=0.34\textwidth]{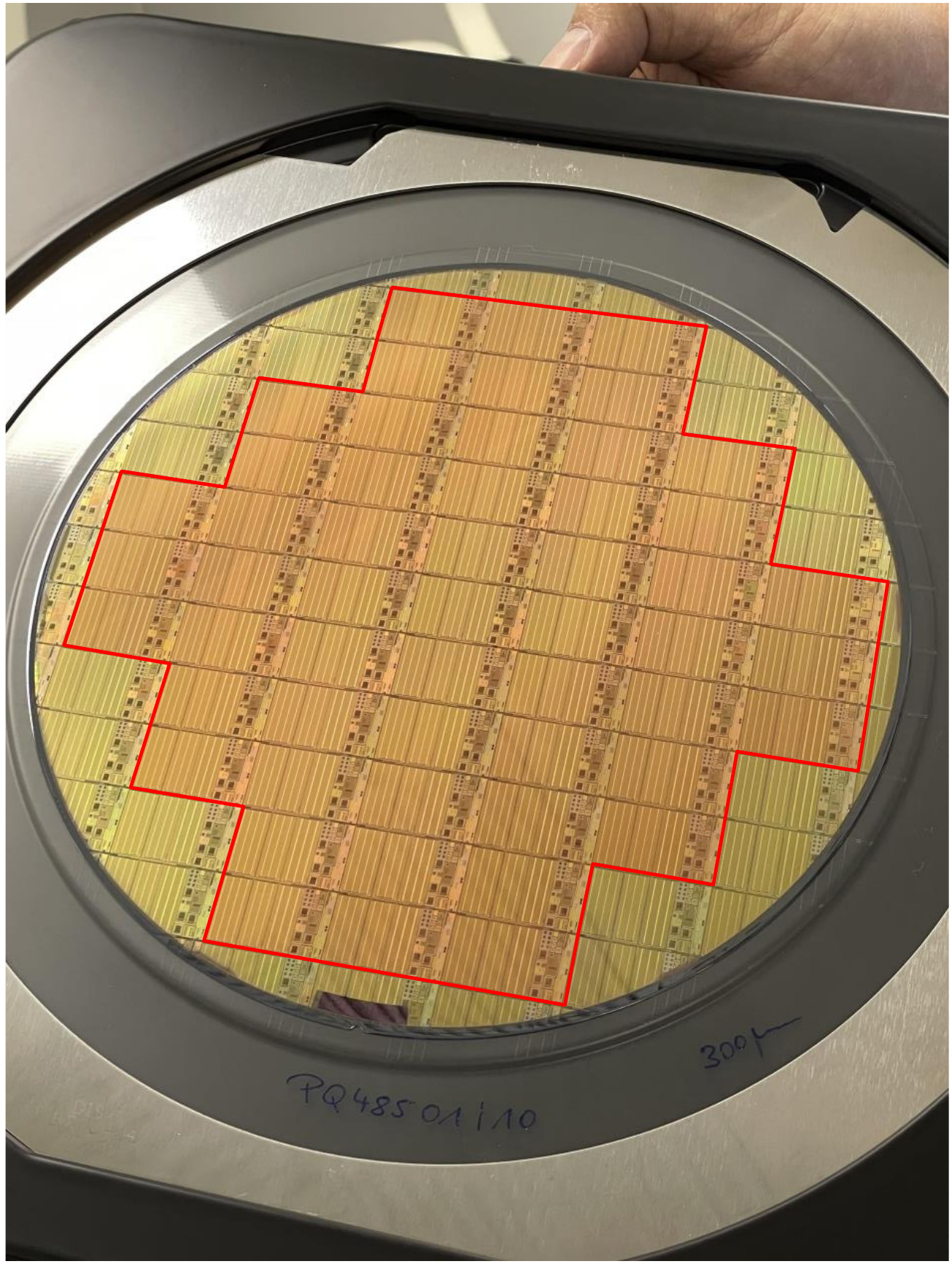}}
\caption{(a) Reticle of the first prototype chip. Several test structures (diodes, TDC, etc), also included in the reticle, are visible on the right side next to the FASER-ALT matrix. (b) Photograph of one of the 8-inch wafers produced. The highlighted region shows the 53 chips available per wafer.}
\label{f:reticle}
\end{figure}

\begin{figure}[ht]
\centering
\subfloat[\label{f:epi}]{\includegraphics[width=0.49\textwidth]{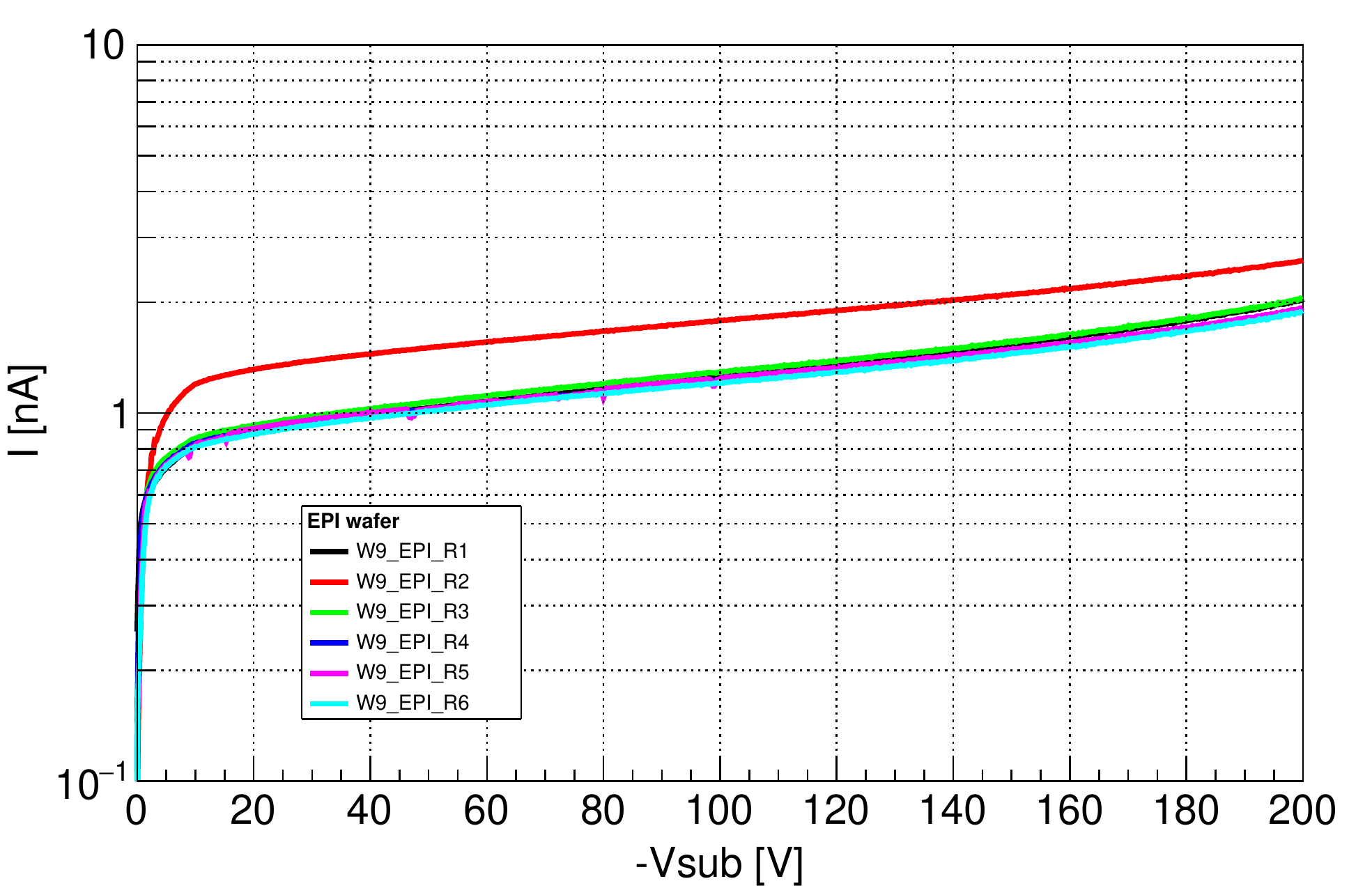}}\hfil
\subfloat[\label{f:noepi}]{\includegraphics[width=0.49\textwidth]{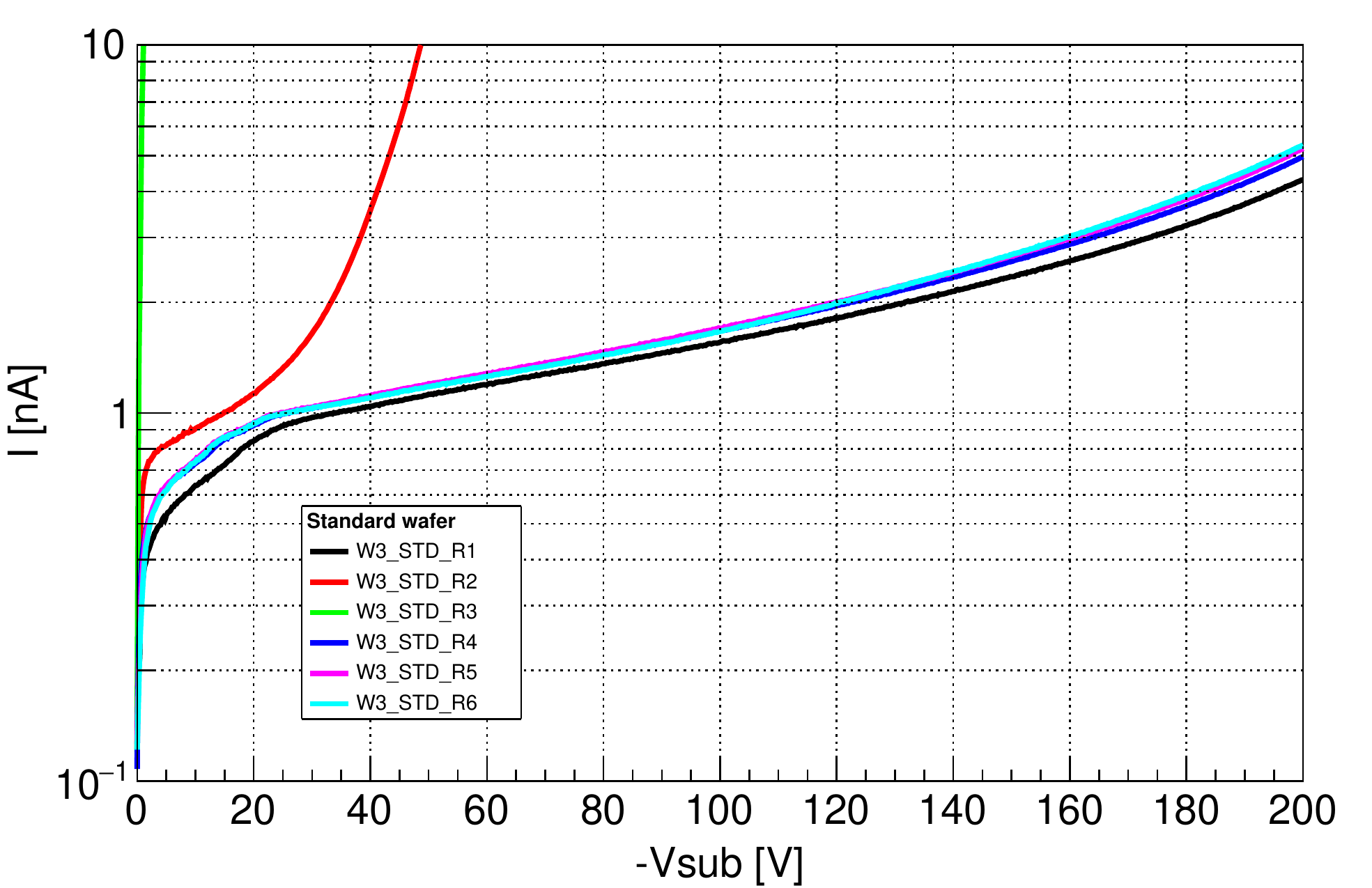}}
\caption{IV curves for chips from wafers produced with a low resistivity substrate and epitaxial layer (a), and with a standard resistivity substrate (b).}
\label{f:iv}
\end{figure}

A first prototype chip has been produced. The chip reticle contains three pixel matrices, so-called FASER-Main, FASER-v2 and FASER-ALT (\cref{f:ret}). FASER-Main contains $128 \times 64$ pixels arranged in four super-columns, with the same design as explained in \cref{sec:chip}. FASER-v2 has $128 \times 48$ pixels (three super-columns), with the same design as FASER-Main excepting the discriminator now being inside each pixel. FASER-ALT contains $128 \times 48$ pixels (three super-columns), without analogue memories and including a counter for charge information. The chips have a total thickness of $\sim \SI{300}{\micro\metre}$ and have been produced by IHP in 8'' (\SI{200}{\milli\metre}) wafers (\cref{f:waf}) on two different substrates: three wafers in a \SI{1}{\ohm\centi\metre} resistivity \SI{230}{\micro\metre}-thick p-type substrate with on top a \SI{350}{\ohm\centi\metre} resistivity \SI{50}{\micro\metre}-thick epitaxial layer, and three other wafers in a standard \SI{50}{\ohm\centi\metre} substrate (without epitaxial layer). The wafer back-side was metallized for the application of the high-voltage to the substrate. Ater dicing, the current-voltage (IV) characteristics of some chips have been measured on a probe-station at the University of Geneva. \Cref{f:iv} shows the IV curves of six chips from a wafer with (\cref{f:epi}) and without (\cref{f:noepi}) epitaxial layer. The measurements were taken at room temperature with the innermost guard-ring connected to ground. The IV-curves are in general very good, with most of the chips reaching \SI{200}{\volt} without breakdown with a leakage current less than \SI{6}{\nano\ampere}.

\begin{figure}[t]
\centering
\includegraphics[width=0.55\textwidth]{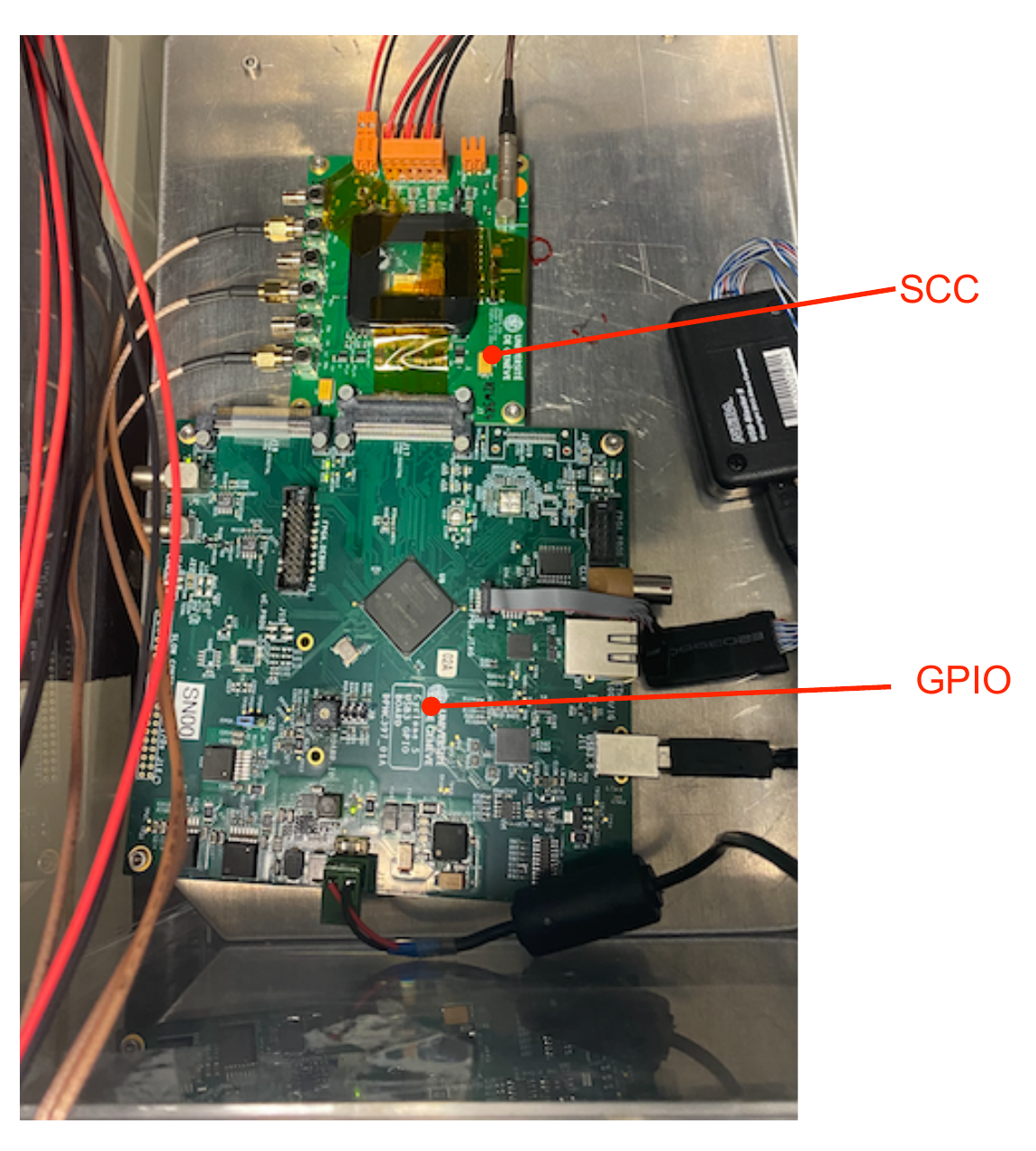}
\caption{\label{f:sccgpio}Single Chip Card (SCC) and GPIO board. The GPIO is based on a Cyclone V A7 FPGA and is interfaced to the control PC via USB3. The SCC is connected to the GPIO via one EMR8 40-pin connector.}
\end{figure}

\begin{figure}[ht]
\centering
\subfloat[\label{f:c0}]{\includegraphics[width=0.48\textwidth]{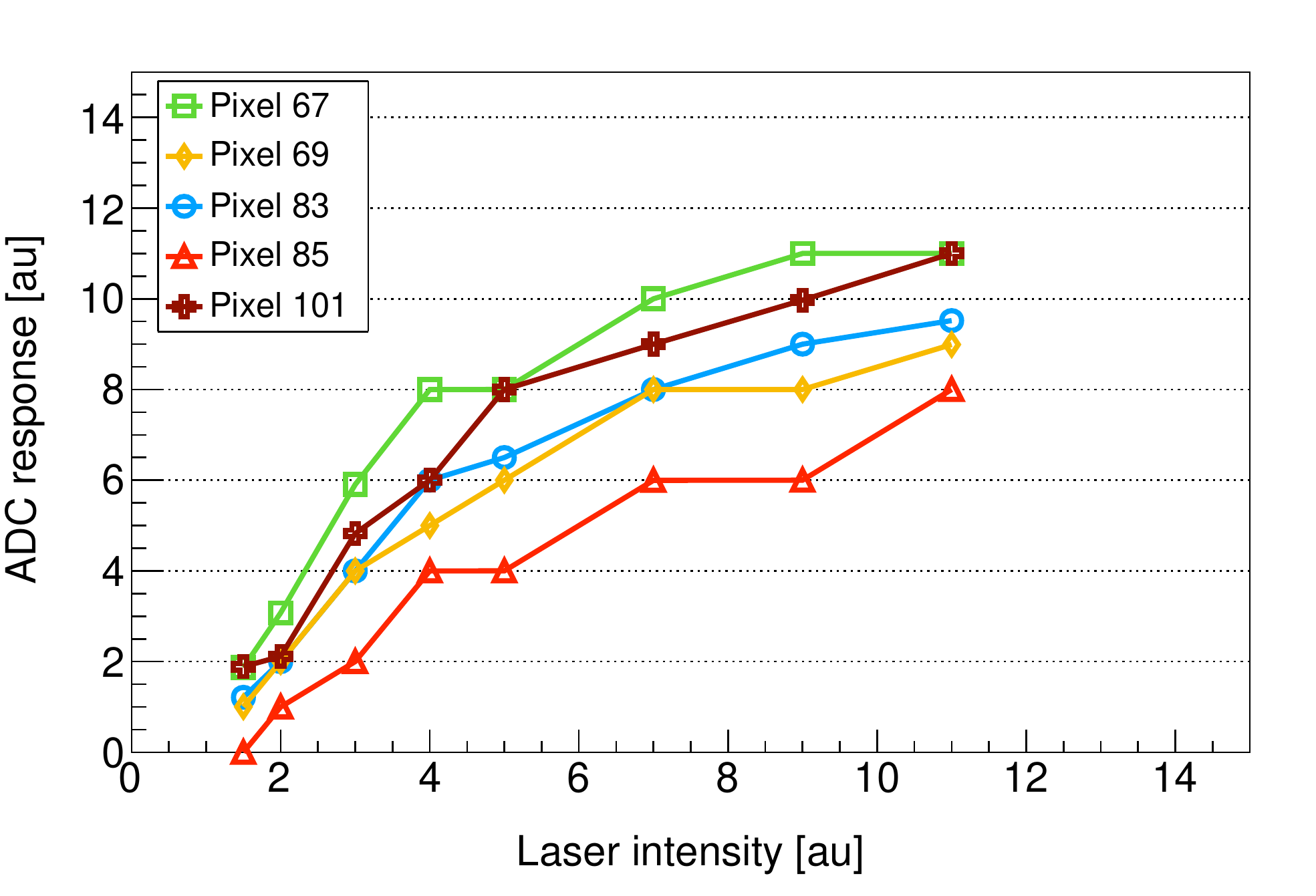}}\hfil
\subfloat[\label{f:c1}]{\includegraphics[width=0.48\textwidth]{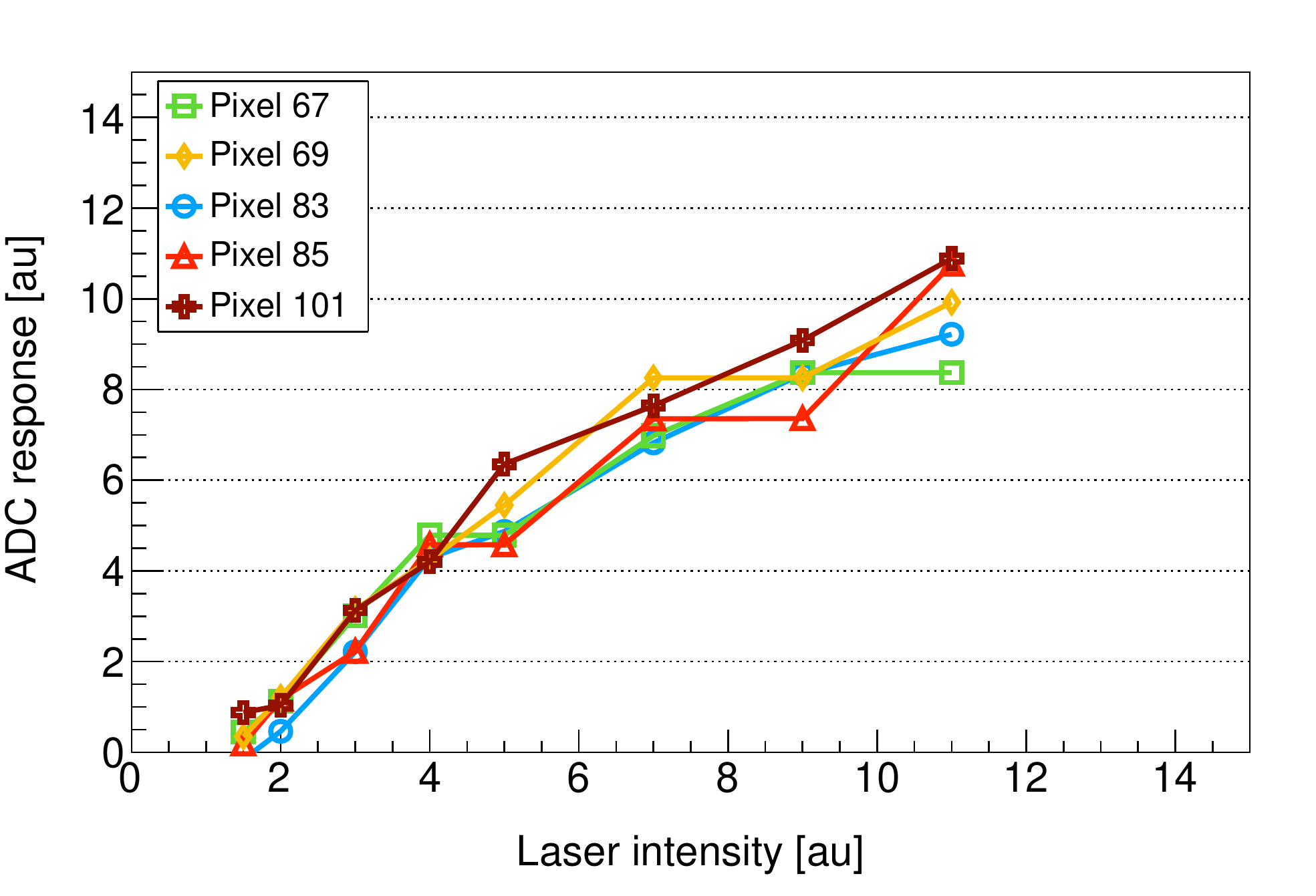}}
\caption{ADC response for different laser intensities before (a) and after (b) calibration using test pulses from the internal injection circuit.}
\label{f:laser}
\end{figure}

Individual pixel matrices have been installed in so-called Single Chip Cards (SCCs), custom PCBs developed to test several chip features. The SCC board includes different independent power domains for the analogue supplies, SMA ports to access the fast-OR output lines and provides access to the guard-rings. The SCC is interfaced to a GPIO board that provides a \SI{10}{\mega\hertz} Serial Peripheral Interface (SPI) for the slow control commands and a \SI{200}{\mega\hertz} clock signal for the data transmission. The readout of the chip has been successfully validated using both the internal test-pulse injection and a \SI{1060}{\nano\metre} infrared laser. \Cref{f:laser} shows the response of the ADC for different laser intensities. In this example, several pixels from a FASER-v2 matrix (substrate with epitaxial layer) with a larger amplifier output mismatch were tested. The laser was focused on selected pixels using three micrometric moving stages. For each laser intensity setting, 2000 pulses were shot and the response of the ADC recorded. The charge is then corrected using calibration curves from the measurement of the ADC response to injected test-pulses of different amplitudes. The resulting spread is largely reduced, these pixels providing now a more uniform response to the same input charge. 

\section{Summary}

A new preshower detector is being developed for the FASER experiment at the LHC. The new preshower will allow to resolve two close-by high energy photons, typically coming from the decay of an axion-like particle produced in the decay volume of the detector. The baseline layout for the new preshower is an array of six pixel detector planes, each composed of twelve modules, interleaved with tungsten absorber layers. The building block of the modules is a new monolithic pixel ASIC developed in 130 nm SiGe BiCMOS technology. The chip has been designed with hexagonal pixels of \SI{65}{\micro\metre} side and a pitch of $\sim \SI{100}{\micro\metre}$. The electronics is contained in a triple-well structure. {A first prototype chip has been produced by IHP  in 8-inch wafers with different substrates. Three different pixel matrices have been included in a $1.5\times 2.4\,\si{\centi\metre\squared}$ reticle. First results indicate good IV characteristics up to 200 V and a functioning readout scheme as tested with the internal charge injection and with an external infrared laser. The detailed characterization of the chip is currently on-going, including the analysis of a testbeam (20 to 120 GeV electron beam) performed at CERN in August 2022 that included several chip prototypes and a FASER calorimeter module. 

\acknowledgments

This work is supported by the Swiss National Science Foundation Grant Nos. 200020\_188489, 200020\_207431 and 20FL21\_201474.


\end{document}